\documentstyle[11pt,newpasp,twoside,epsf]{article}
\def\edcomment#1{\iffalse\marginpar{\raggedright\sl#1\/}\else\relax\fi}
\reversemarginpar

\begin{document}
\title{
Modern Techniques in Galaxy Kinematics: Results from Planetary Nebula Spectroscopy
}
\author{Aaron J. Romanowsky$^1$, Nigel G. Douglas$^1$, Konrad Kuijken$^1$, Magda Arnaboldi$^2$, Joris Gerssen$^3$, \& Michael R. Merrifield$^4$}
\affil{{}$^1$Kapteyn Institute, Postbus 800, 9700 AV Groningen, The Netherlands
{}$^2$Osservatorio di Capodimonte, Via Moiariello 16, I-80131 Naples, Italy
{}$^4$STScI, 3700 San Martin Drive, Baltimore, MD 21218, USA \\
{}$^3$School of Physics \& Astronomy, Univ. Nottingham, NG72RD, England
}

\begin{abstract}
We have observed planetary nebulae (PNe) in
several early-type galaxies
using new techniques on 4- to 8-meter-class telescopes.
We obtain the first large data sets ($\ga 100$ velocities each)
of PN kinematics in galaxies at $\ga 15$ Mpc,
and present some preliminary dynamical results.
%The accompanying poster (Douglas et al. 2002) introduces the Planetary Nebula Spectrograph (PN.S), a dedicated instrument for tracing the kinematics of stars in external galaxies.
%This is an important part of a multi-pronged approach to collecting planetary nebula kinematical data in a large sample of early-type galaxies to a distance of 20 Mpc. The
%methods employed include multi-fiber spectroscopy with WYFFOS+AUTOFIB2 on the 4.2-meter William Herschel Telescope (WHT); counter-dispersed imaging with ISIS
%and the PN.S on the WHT; and "masked counter-dispersed imaging" with FORS2+MXU on the ESO Very Large Telescope. We present the fruits of these methods: ~100-200
%PN velocities in each of the elliptical galaxies NGC 821, NGC 4472, NGC 4486, NGC 5866, and NGC 7457, from which we can estimate the galaxies' mass and rotational
%properties using simple models. We also describe results from ongoing rigorous dynamical analyses using orbit modeling methods developed for discrete velocity data
%(Romanowsky \& Kochanek 2001). 
\end{abstract}

Planetary nebulae (PNe) have great promise as dynamical tracers 
of the diffuse stellar population in external galaxies.
Especially important is their use in the outer parts of early-type galaxies
(ellipticals and S0s), an arena which is not generally accessible with other
dynamical tracers.
But acquisition of PN velocities 
in galaxies farther away than a few Mpc has been quite limited.
We report here results from new observing programs 
of early-types at
$\ga 15$ Mpc.

The standard method for acquiring PN velocities involves a narrow-band
imaging survey followed by multi-object spectroscopy (MOS).
PN MOS in galaxies at $\ga 15$ Mpc
has been not very successful in practice, largely owing to
astrometric difficulties.
To circumvent this
we have developed
{\it masked counter-dispersed imaging},
a technique using a standard multi-slit mask with enlarged (4\arcsec) slits
to absorb any astrometry errors;
we take two images dispersed in opposite directions to
permit the velocity determination.
Using this technique with FORS2/MXU at UT2 on 28-30 April 2001, we have obtained 
PN velocities to 4~$R_{\rm eff}$ (effective radii)
in the Virgo Cluster giant ellipticals
M87 = NGC 4486 (200 PNe; see Fig. 1) 
and M49 = NGC 4472 (80 PNe, including some obtained with
the multi-fiber unit WYFFOS/AF2 at the WHT on 25-28 May 2000).

In the outer regions of M87, the PNe 
rotate about the minor axis, with lower velocities than
the globular clusters (GCs) 
($\sim 100$ km~s$^{-1}$ vs. 200-300 km~s$^{-1}$;
C\^{o}t\'{e} et al. 2001).
The inferred stellar velocity dispersion $\sigma_{\rm p}(R)$ rises slightly with radius,
as seen in the GCs but with lower amplitude.
Simple Jeans models suggest the stars and GCs have 
significantly different dynamics.
In M49, the stars and GCs (Zepf et al. 2000) 
show more consistent kinematics,
with $\sigma_{\rm p}(R)$ constant or slightly rising with radius.

The most efficient technique for finding and measuring velocities of 
extragalactic PNe is {\it counter-dispersed imaging} (see
the accompanying paper, Douglas et al. 2002).
Using the WHT's standard spectrograph ISIS in slitless mode
in April 1997 and June 1998, we have
obtained 90 PN velocities to 4~$R_{\rm eff}$ in the S0 galaxy NGC 5866.
With the newly commissioned PN.Spectrograph (PN.S) at the WHT
on 13-18 September 2001,
we have obtained ~100 PN velocities to 5~$R_{\rm eff}$
in each of the galaxies NGC 821 (E3)
and NGC 7457 (S0).
NGC 5866 and NGC 7457 show rotational dominance at large radii.
In NGC 821, $\sigma_{\rm p}(R)$ declines at large radii,
contrary to what would be expected for a round giant elliptical from
more radially limited dynamical studies
(Gerhard et al. 2001),
but similar to what is seen in the E3 galaxy NGC 4697
(M\'{e}ndez et al. 2001).

The PN.S will embark in March 2002
on a kinematical survey of a dozen nearby bright ellipticals.
Key interests in all these PN studies of early-type galaxies  
are the amount of angular momentum in their outer parts,
their distribution of mass, 
and tracers of their formational processes via relict kinematical substructure.
Rigorous analysis of the data will proceed with 
nonparametric orbit modeling methods
designed to include discrete velocity measurements
(Romanowsky \& Kochanek 2001).

\begin{figure}
\plottwo{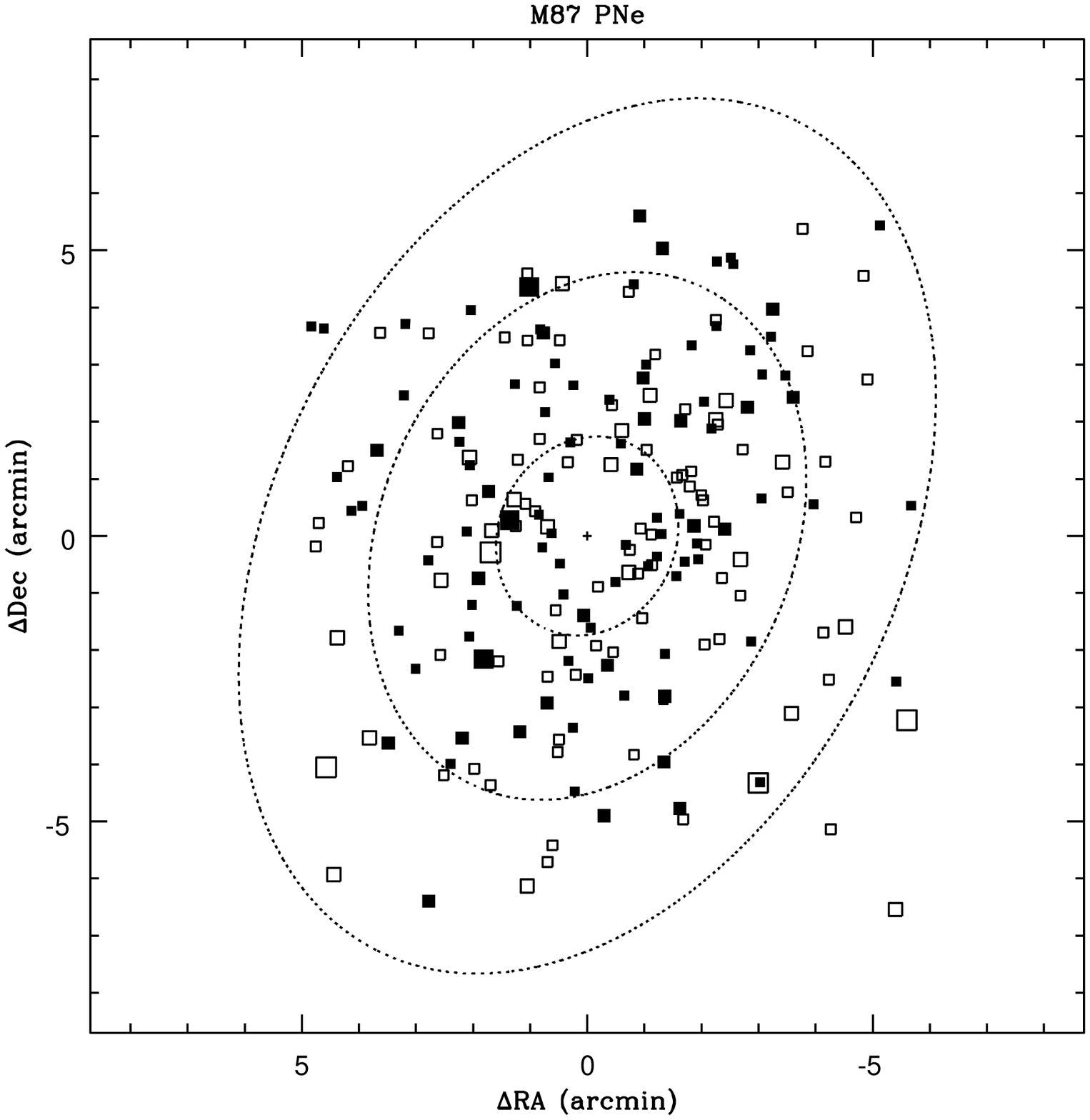}{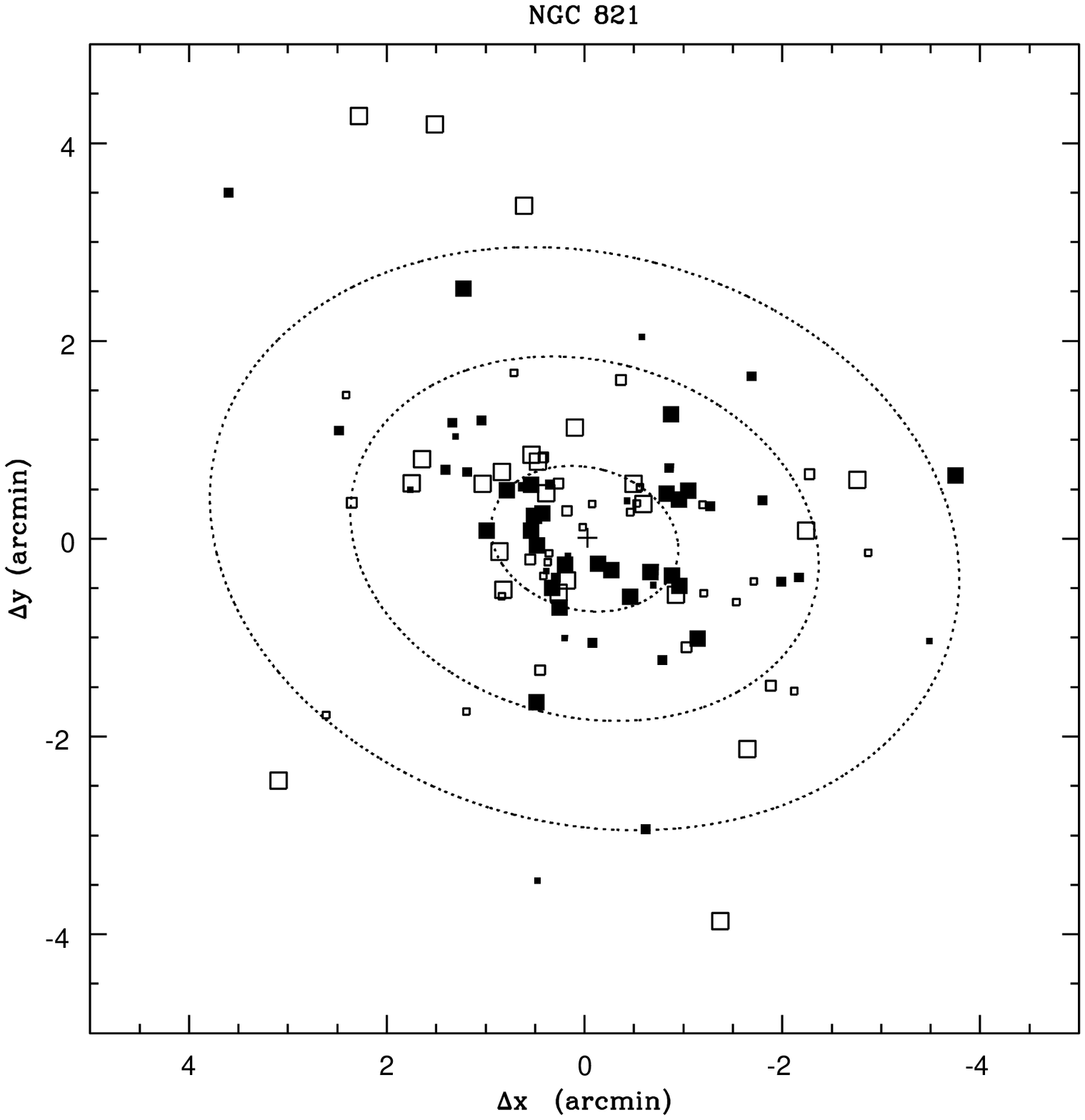}
\caption{PN velocities in M87 ({\it left}, 200 PNe at 15 Mpc, using UT2+FORS2/MXU) 
and in NGC 821 ({\it right}, 104 PNe at 25 Mpc, using WHT+PN.S).
Filled and open boxes represent approaching and receding velocities,
respectively, relative to the systemic velocity;
the box size indicates the relative velocity magnitude.
Dotted lines indicate the stellar isophotes at 1~$R_{\rm eff}$,
2.5~$R_{\rm eff}$, and 4~$R_{\rm eff}$.
}
\end{figure}

\end{document}